# Structural characterization of Nd-doped calcium aluminosilicate glasses designed for the preparation of zirconolite (CaZrTi$_2$O$_7$) -based glass-ceramic

P. Loiseau, D. Caurant, N. Baffier

*CNRS, Ecole Nationale Supérieure de Chimie de Paris (ENSCP, Paristech), Laboratoire de Chimie de la Matière Condensée de Paris (UMR 7574), 11 rue Pierre et Marie Curie, F-75005 Paris, France*

K. Dardenne, J. Rothe, M. Denecke

*FZK GmbH, INE, Hermann-von-Helmoltz-Platz, D-76344 Eggenstein-Leopoldshafen, Germany*

S. Mangold

*FZK GmbH, ISS, Hermann-von-Helmoltz-Platz, D-76344 Eggenstein-Leopoldshafen, Germany*

*This work concerns glasses belonging to SiO$_2$-Al$_2$O$_3$-CaO-ZrO$_2$-TiO$_2$-Nd$_2$O$_3$ system that lead to zirconolite crystallization in their bulk after nucleation + crystal growth thermal treatments and that could find application as nuclear waste form. The understanding of crystallization processes in glasses implied to investigate their structure. The environment around Ti, Zr (nucleating agents) and Nd was characterized for various Nd$_2$O$_3$ loadings. Electron spin resonance study of the small amount of Ti$^{3+}$ occurring in glasses enabled to identify two types of sites for titanium. EXAFS showed that Zr occupied a quite well defined 6-7-fold site coordinated by oxygen at 2.20 Å, with second neighbors that could correspond to Ca/Ti and Zr around 3.5 Å. This short range order presents some similarities with zirconolite, which could predispose glass to zirconolite nucleation. Nd environment was probed by optical spectroscopies, ESR and EXAFS. Results showed that the environment around Nd was very constrained by the glassy network and significantly differed from the one in zirconolite. Nd occupied a highly distorted 8-9-fold coordinated site in glass with oxygen atoms at 2.53 Å. No second neighbors were clearly identified by EXAFS around Nd, but the study of Nd optical fluorescence decays suggested a strong interaction between Nd ions.*

## INTRODUCTION

Zirconolite (CaZrTi$_2$O$_7$) is a naturally occurring phase which exhibits an excellent chemical durability and a good actinide containment capacity. This is a waste form of choice for the specific immobilisation of minor actinides or plutonium. A simpler alternative to the ceramic route is to prepare zirconolite crystals by controlled crystallization (nucleation+growth) of a parent glass. In addition, such glass-ceramic offers a greater chemical flexibility against waste composition fluctuations due to the presence of a residual glass surrounding zirconolite crystals. The basic parent glass composition which was studied is (mol. %): 48.8 SiO$_2$, 8.5 Al$_2$O$_3$, 25.3 CaO, 11.3 TiO$_2$, 5.0 ZrO$_2$, 1.1 Na$_2$O[1]. Various loadings of Nd$_2$O$_3$ (trivalent actinide surrogate) were added up to 2.2 mol. % (10 wt. %). Such glasses led to the only crystallization of zirconolite by devitrification (nucleation+crystal growth)[2]. To get insight into crystallization processes, the environment around three elements was investigated: Ti and Zr (main components of zirconolite playing a key role in nucleation processes), and Nd.

**EXPERIMENTAL**

40g batch glasses were prepared following a standard method comprising two stages of melting and casting at 1550°C with intermediate grinding. A similar glass transformation temperature around 760 °C was found for all the glasses, so that the addition of $Nd_2O_3$ up to 2.2 mol. % has no significant effect on the mean bond strength of the glassy network.

$Ti^{3+}$ Electron Spin Resonance (ESR) measurements were performed at 120 K using a Bruker ESP 300e spectrometer operating at X-band. Room temperature EXAFS (Extended X-ray Absorption Fine Structure) measurements were performed at Zr-K edge (17998 eV) and Nd-$L_3$ edge (6208 eV) at ANKA synchrotron (Germany, Karlsruhe) in transmission mode through quenched glass samples containing 1.3 mol. % of $Nd_2O_3$ (6 wt. %). EXAFS signals were simulated using the UWXAFS package[3]. $Nd^{3+}$ optical properties were characterised at 15 K by absorption from 400 to 900 nm and by emission from 860 to 960 nm ($^4F_{3/2} \rightarrow {}^4I_{9/2}$ channel). At 15 K, only the lowest energy Kramers doublet of the $^4I_{9/2}$ or of the $^4F_{3/2}$ multiplet is populated for absorption or emission measurements respectively.

**RESULTS AND DISCUSSION**

### ESR study of native $Ti^{3+}$ ions in parent glass

During glass meting, a $Ti^{4+} \Leftrightarrow Ti^{3+}$ equilibrium was established and then was frozen by quenching at room temperature. Therefore, a small quantity of $Ti^{3+}$ formed. Their amount was quantified by ESR at 120 K (Fig. 1) with a DPPH standard: the proportion of Ti at the trivalent oxidation state was below 0.7 ppm (case of the undoped glass) and decreased with increasing $Nd_2O_3$ concentration (no $Ti^{3+}$ ion detected for the 10 wt. % $Nd_2O_3$ glass). This evolution could result from an increase of the glass basicity[4] with Nd concentration, if we admit that Nd acts as a network modifier.

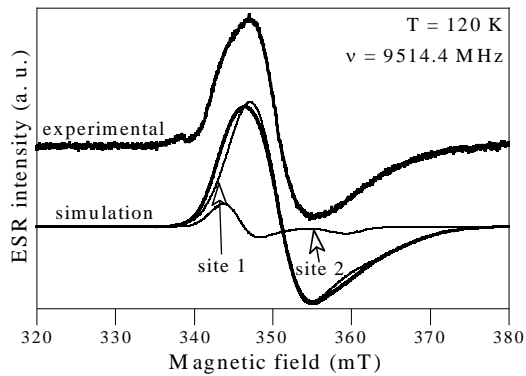

Fig. 1. Experimental and simulated $Ti^{3+}$ ESR signal for the undoped glass

$Ti^{3+}$ ESR signal was simulated with the assumption of an axial symmetry for simplicity reasons. Without any distribution of g factor, it was necessary to consider two contributions (with gaussian line shape) to simulate the signal and the shoulder occurring around 344 mT

(Fig. 1 and Table I). This gives evidence of two types of sites for $Ti^{3+}$ ions in the glass, referred as site 1 and site 2. Site 2 contributes only to 8.5 % of the overall signal.

|        | $g_\perp$ | FWHM   | $g_{//}$ | FWHM  |
|--------|-----------|--------|----------|-------|
| Site 1 | 1.946     | 6.5 mT | 1.88     | 12 mT |
| Site 2 | 1.973     | 4 mT   | 1.89     | 4 mT  |

TABLE I. Results of the ESR simulation.

The g factors give pieces of information about the symmetry of $Ti^{3+}$ environment in the glass[5]. As $Ti^{3+}$ ESR signal can be detected at room temperature, several simple geometries can be excluded: tetrahedron, octahedron and tetragonally elongated octahedron. Moreover, a four-fold coordination is quite unlikely for a cation as big as $Ti^{3+}$ ($r_{VI}(Ti^{3+})$=0.67 Å> $r_{VI}(Ti^{4+})$=0.605 Å[6]). However, ESR results can be fully explained by considering a compressed octahedral symmetry, highly distorted to have the order $g_{//} < g_\perp$ (more distorted for site 2). This geometry could correspond either to a $C_{4v}$ or a $D_{4h}$ symmetry (5 to 6 coordination number). It is interesting to notice that the $C_{4v}$ symmetry is the one of the square pyramidal $O=TiO_4$ frequently proposed for $Ti^{4+}$ environment in glasses[7].

### EXAFS characterisation of $Zr^{4+}$ environment in parent glass

The short range order around Zr in the 6 wt. % $Nd_2O_3$ parent glass was determined from the EXAFS spectrum (Fig. 2). Zr occupies a well defined site (debye-waller factor $\sigma^2 = 0.007 \pm 0.001$ Å$^{-2}$) of 6.5 ± 0.7 coordination number with a mean Zr-O distance d(Zr-O) = 2.15 ± 0.02 Å. This environment is relatively close to the one in a depolymerised glass such as R7T7 glass, with a coordination number around 6 and a mean Zr-O distance ranging from 2.07 to 2.10 Å[8]. The slightly higher coordination number and Zr-O distance in our glass could be due to the higher polymerisation of the glassy network. For 100 moles of undoped glass, assuming that Al, Ti and Zr adopt respectively the geometries $(AlO_4)^-$, $(O=TiO_4)^{2-}$ and $(ZrO_6)^{2-}$, 24.7 moles of $(CaO+Na_2O)$ would be mobilised for their charge compensation, so that only 1.7 moles of $(CaO+Na_2O)$ would be free to depolymerize the glassy network. Then, oxygens are less polarisable and a higher coordination number is required for $Zr^{4+}$ charge shielding. In zirconolite, Zr coordination number is 7 and d(Zr-O) = 2.20(1) Å. This short range order similarity with the parent glass could predispose to zirconolite nucleation.

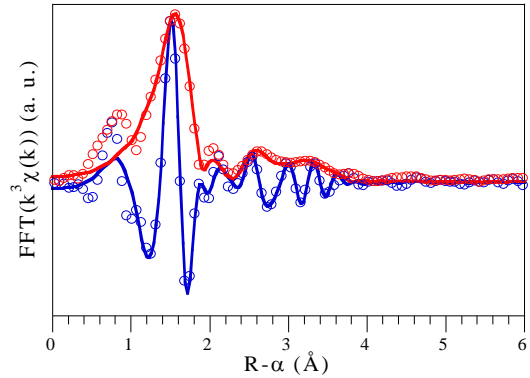

Fig. 2. Fit (solid lines) to the EXAFS (open circles) of the 6 wt. % $Nd_2O_3$ parent glass (Zr-K edge, room temperature).

The EXAFS analysis of the next nearest neighbours around Zr is complicated (Fig. 2): at least two shells (likely three) with overlapping shapes are necessary for fitting, and they partly interfere destructively. Nevertheless, the Fourier peak around 3.2 Å seems to be associated to Ca/Ti on the one hand and Zr on the other hand, both occurring around 3.5 Å from Zr.

**Spectroscopic investigation of $Nd^{3+}$ environment in parent glass**

The fluorescence spectra corresponding to the $^4F_{3/2} \rightarrow \ ^4I_{9/2}$ Nd transition were recorded at 15 K by exciting the Nd-doped glasses around 806.5 nm. These spectra allow to estimate the total splitting of the $^4I_{9/2}$ ground multiplet which is about 500 cm$^{-1}$, against 775 cm$^{-1}$ for Nd-doped zirconolite samples. The $^4I_{9/2}$ splitting is a measure of the mean crystal field experienced by $Nd^{3+}$ ion in a host[9]. The greater crystal field of Nd in zirconolite as compared to parent glass indicates not only that the environments are different between the two hosts, but also that the mean Nd-O distance is shorter in zirconolite.

Nd optical absorption spectra were recorded for the Nd-doped glasses at 15 K. They are all very similar, even though a slight broadening of absorption lines occurs with increasing $Nd_2O_3$ concentration from 0.5 to 10 wt. %. Three absorption bands from $^4I_{9/2}$ state are particularly interesting: towards $^2P_{1/2}$ (~ 23200 cm$^{-1}$), $^4F_{3/2}$ (~ 11400 cm$^{-1}$) and $^2G_{7/2}$ - $^4G_{5/2}$ (~ 17200 cm$^{-1}$). The first two bands ($^2P_{1/2}$ and $^4F_{3/2}$) notably give information about the variety of Nd environment because of the low degeneracy of the excited state, whereas the latter band ($^2G_{7/2}$ - $^4G_{5/2}$) is described as hypersensitive and is very sensitive to Nd environment. For instance, the FWHM of the $^4I_{9/2} \rightarrow \ ^2P_{1/2}$ band is approximately 150 cm$^{-1}$ in Nd-doped parent glasses, much more than the one measured for zirconolite (30 cm$^{-1}$), which reflects a higher distribution of environment in the glasses. Moreover, by comparison with literature[10], the shape of the $^4I_{9/2} \rightarrow \ ^2G_{7/2}$ - $^4G_{5/2}$ is structure-less and is characteristic either of highly polymerised glassy network or of glasses containing high field strength network modifiers. All these observations agree with an irregular Nd environment partly imposed by the glassy network topology, contrarily to Zr.

The environment of Nd was probed by EXAFS at $L_3$ edge (Fig. 3). Nd occupies a highly distorted site: the EXAFS simulation implies a cumulant analysis of the data (non-gaussian asymmetric Nd-O pair distribution[11]) with a large $\sigma^2 = 0.029 \pm 0.001$ Å$^{-2}$. Nd-O coordination number is 8.6 ± 0.8 and the mean Nd-O distance is about 2.53 ± 0.02 Å in the glass.

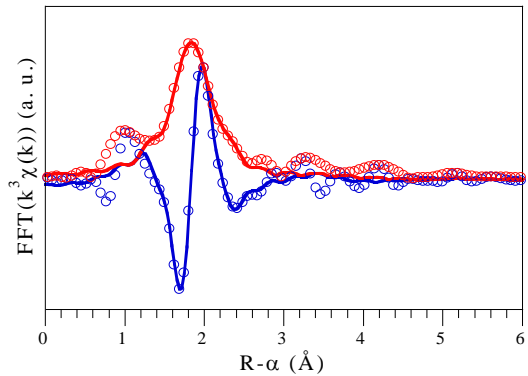

Fig. 3. Fit (solid lines) to the EXAFS (open circles) of the 6 wt. % $Nd_2O_3$ parent glass (Nd-$L_3$ edge, room temperature).

In zirconolite, Nd coordination by O is 8 and, by EXAFS, Nd-O bond distance was found to be 2.44±0.01 Å with $\sigma^2$=0.010±0.001 Å$^{-2}$. So, the environment around neodymium in the glass shows striking differences with the one in zirconolite and is much more distorted. Therefore, the incorporation of neodymium into zirconolite crystals during devitrification will imply strong reconstructive phenomena which are assumed to occur at high temperature, when the viscosity of the supercooled liquid is lowered. In these conditions, neodymium can hinder zirconolite nucleation with increasing diffusionnal difficulties.

## CONCLUSION

ESR study of Ti$^{3+}$ ions formed during glass melting revealed that their environment could correspond to a square pyramidal O=TiO$_4$ or to a six-fold coordinated octahedron axially compressed. The coordination numbers involved in such geometries are close to the ones encountered in zirconolite (6 and 5) which is aimed to crystallize by devitrification. Zr-K edge EXAFS spectra showed that Zr$^{4+}$ ions occupied well defined 6-7 coordinated sites, close to the ones of zirconolite, able to explain the nucleating effect of Zr in the glass. Contrarily to Zr$^{4+}$ ion which can be described as a cross-linking agent that reinforces the glassy network, Nd$^{3+}$ ion must rather be considered as a network modifier or a charge compensator which exhibits a lower ability to impose its own environment. All the optical spectroscopies and EXAFS agree to describe Nd environment as being constrained by the glassy network with coordination number and Nd-O distance significantly greater than the ones of zirconolite: Nd$^{3+}$ ion cannot act as a nucleating agent for zirconolite crystallization.